\documentclass[authoryear]{elsarticle}
\usepackage[utf8]{inputenc}
\usepackage[margin=3cm]{geometry}

\usepackage{amsmath,amssymb,amsfonts,amsthm}
\usepackage{siunitx}  

\usepackage{graphicx}
\usepackage{subcaption}  
\usepackage[section]{placeins}  

\usepackage[flushleft]{threeparttable}  
\usepackage{booktabs}  
\usepackage{multirow}  
\newcolumntype{H}{>{\setbox0=\hbox\bgroup}c<{\egroup}@{}}  

\usepackage{acronym}

\newcommand{\degree}{^{\circ}}

\newcommand{\vecw}{\vec{w}}

\begin{document}

\begin{frontmatter}
\title{Hybrid Search method for Zermelo's navigation problem}

\author[uca,cg]{Daniel Precioso}\ead{daniel.precioso@canonicalgreen.com}
\author[dal]{Robert Milson}\ead{rmilson@dal.ca}
\author[dal]{Louis Bu}\ead{l.bu@dal.ca}
\author[dal]{Yvonne Menchions}\ead{vg217777@dal.ca}
\author[ie,cg]{David Gómez-Ullate\corref{corauthor}}\ead{dgomezullate@faculty.ie.edu}

\cortext[corauthor]{Corresponding author}
\affiliation[uca]{organization={Department of Computer Science, Higher School of Engineering, Universidad de Cádiz},
            addressline={Av. Universidad de Cádiz, 10},
            postcode={11519},
            city={Puerto Real, Cádiz},
            country={Spain}}
\affiliation[cg]{organization={Canonical Green},
            city={Madrid, Madrid},
            country={Spain}}
\affiliation[dal]{organization={Department of Mathematics and Statistics, Dalhousie University},
            city={Halifax, Nova Scotia},
            postcode={B3H 3J5},
            country={Canada}}
\affiliation[ie]{organization={School of Science and Technology, IE University},
            addressline={Paseo de la Castellana, 259},
            postcode={28046},
            city={Madrid, Madrid},
            country={Spain}}

\begin{abstract}
In this paper, we present a novel algorithm called the \acl{HS} algorithm to tackle the \acl{ZNP}. This method can be regarded as an extension of the recent \acl{FMS} to allow for further exploration in search for the global optimum, in situations of complex vector fields where many locally optimal trajectories exist. Our algorithm is designed to work in both Euclidean and spherical spaces and utilizes a heuristic that allows the vessel to move forward while remaining within a predetermined search cone centred around the destination. This approach not only improves efficiency but also includes obstacle avoidance, making it well-suited for real-world applications. We evaluate the performance of the \acl{HS} algorithm on synthetic vector fields and real ocean currents, demonstrating its effectiveness and performance.
\end{abstract}


\end{frontmatter}


\section*{Acronyms}
\begin{acronym}
    \acro{FMS}{Ferraro-Martín de Diego-Sato algorithm}
    \acro{HS}{Hybrid Search}
    \acro{RK4}{Fourth order Runge-Kutta method}
    \acro{ZNP}{Zermelo's navigation problem}
    \acro{ZIVP}{Zermelo's initial value problem}
\end{acronym}


\section{Introduction}
\label{sec:intro}

Weather routing is the process of using mathematical models and weather forecasting data to determine the most efficient route for a vessel to take \cite{Shao2012}. This is done by considering factors such as the wind and current conditions, as well as the vessel's characteristics and constraints \cite{Zhou2019}, in order to find the route that will require the least amount of fuel and time to complete. From a mathematical perspective, weather routing involves solving optimization problems in order to find the optimal route \cite{Lin2013a, Walther2016}. This typically involves using algorithms such as dynamic programming \cite{Klompstra1992, Wei2012} or graph search \cite{Gkerekos2020, Shin2020} to find the best combination of route and speed that will minimize the time and fuel consumption of the vessel.

A classic formulation for weather routing in the field of mathematics is the \acf{ZNP} on the plane \cite{Zermelo1930}. Formally, \ac{ZNP} involves optimizing the time it takes for a ship to travel through a space while being affected by a perturbation, which can be a vector field representing a wind or a current. Zermelo himself solved this problem in $\mathbb{R}^2$ and $\mathbb{R}^3$ \cite{Zermelo1931}, and was generalized shortly after by \citet{Levi1931} to an $n$-dimensional plane $\mathbb{R}^n$ and solved by means of variational calculus. Research on \ac{ZNP} is still ongoing,  following investigation on its generalizations \cite{Kopacz2019, Shavakh2022} and analysis on the effect of singularities \cite{Bonnard2021, Bonnard2022}.

To make the \ac{ZNP} applicable in real-world scenarios, researchers have developed more realistic formulations that focus on modelling Earth as a sphere using $\mathbb{S}^2$ \cite{Bao2004, Marchidan2016}. Other lines of research have explored time-dependent vector fields \cite{Aldea2020, Aldea2021} and obstacle avoidance \cite{Li2013, Lai2021}.

In this paper, we propose a novel \acf{HS} algorithm that combines the \ac{ZNP} with the \acf{FMS} algorithm \cite{Ferraro2021, Ferraro2022} to find the optimal path for a ship to reach its destination. Our approach is applicable both in Euclidean and spherical spaces. Unlike traditional isochrones, which involve sending a ship for a set amount of time and identifying all the possible locations it could reach in that time, our approach utilizes a heuristic that allows the ship to move forward as long as it heads in the right direction (i.e., remains within a predetermined search cone centred around the destination). This method improves efficiency and automatically includes obstacle avoidance, making it well-suited for real-world applications. Careful tuning of the free parameters in \acf{HS} allows to balance exploration and exploitation, thus reaching a compromise between computational efficiency and the quality of the solution.  

The paper is structured as follows. Section \ref{sec:problem} introduces the routing problem and the \ac{ZNP} equations. In section \ref{sec:hybrid}, we use these equations to introduce our \ac{HS} method. Section \ref{sec:benchmarks} introduces synthetic and real benchmarks, and section \ref{sec:results} presents the results of the \ac{HS} algorithm on these benchmarks. Finally, we discuss the results in section \ref{sec:discussion}.


\section{Routing problem in navigation}
\label{sec:problem}

The routing problem refers to the challenge of identifying the most efficient and safe route for ships or vessels to travel from one point to another. The routing problem is complex, as it depends on several factors, such as weather conditions, ocean currents, vessel characteristics, and safety concerns. Navigation routing problems have practical implications, as finding an optimal route can reduce fuel consumption, emissions, and operating costs, while also improving safety and reducing travel time. The routing problem is highly complex, and it is not possible to find an optimal route manually. However, the problem can be addressed with the help of mathematics and computation. In this paper, we propose a method to find time optimal routes taking into account the effect of ocean currents.

We are taking several assumptions to simplify the problem. First, although it is known that weather conditions change through time, we are assuming a stationary vector field for the sake of simplicity, though the method proposed in this study is easy to adapt for evolving conditions. Second, we assume the ship keeps a constant velocity with respect to water. Lastly, real case scenarios have obstacles present, in the form of land. Our method is able to avoid small obstacles but it is not intended to perform optimal circumnavigation, i.e. bypass significant obstacles in the way to the goal.

To address the problem, we will use the classical formulation from \acf{ZNP}. This section introduces the equations both on the plane and on the sphere.

\subsection{Zermelo's Navigation Problem on the plane}
\label{sec:zermelo-euclidean}

This problem was proposed in 1931 by Ernst Zermelo \cite{Zermelo1931}, is a classic time-optimal control problem, where its aim is to find \textbf{time minimum trajectories} under the influence of a drift vector
\[\vecw(x_1,x_2) = \left< w_1(x_1,x_2), w_2(x_1,x_2)\right>\]
where $x_1,x_2$ are local coordinates, and where $w_1, w_2$ are the vector components chosen relative to a local frame. This drift vector can be interpreted as wind or water current. In small scale simulations, the coordinates and the vector components can be taken to be Euclidean. Once we pass to larger scale simulations that take into account the Earths' curvature, the coordinates $(x_1,x_2)$ indicate longitude and latitude delineated in degrees, while the vector components are taken relative to a local east-north framing and delineated in meters.

The goal is to navigate from a specified initial point along a path that minimizes time, under the influence of $\vecw$, assuming the vessel provides constant thrust $V$ (speed over water) and has a heading angle (over water) $\alpha$ w.r.t. the $x_1$-axis. Thus, the velocity components over ground can be expressed as:

\begin{align}
\begin{split}
  \frac{d x_1}{d t} & = V \cos \alpha + w_1(x_1,x_2) \\
  \frac{d x_2}{d t} & = V \sin \alpha +
  w_2(x_1,x_2)
\end{split}
\label{eq:dxdt}
\end{align}

Using the Calculus of Variations, one can show that such a path necessarily obeys the following differential equation, first derived by Zermelo \cite{Zermelo1931}
\begin{equation}
  \frac{d\alpha}{dt} = \sin^2(\alpha)\, w_{2,1} + \sin(\alpha) \cos
 (\alpha)\, \left( w_{1,1}-w_{2,2}\right) - \cos^2(\alpha)\,   w_{1,2}
\label{eq:dthetadt}
\end{equation}
where for the sake of brevity, we write $w_{i,j} = \partial w_i / \partial x_j$. 

Equation \eqref{eq:dthetadt} is known as the \textbf{Zermelo differential equation}.  Together with \eqref{eq:dxdt} it gives the form for time-optimal trajectories as a dynamical system in the 3-dimensional space parameterized by $(x_1, x_2, \alpha)$. We will refer to the initial value problem for this 3-dimensional dynamical system as the \ac{ZIVP}. This means that given a current vector field $(w_1(x_1,x_2),w_2(x_1,x_2))$, an initial position $(x_1^{(0)},x_2^{(0)})$ and an initial heading $\alpha$, the trajectory defined by the initial value problem is guaranteed to be time optimal, i.e. each point in that trajectory cannot be reached in shorter time by a vessel with constant speed over water $V$ starting from $(x_1^{(0)},x_2^{(0)})$.
For the interest of completeness, the derivation of this last equation is fully explained in Appendix \ref{appendix:zermelo}.

\subsection{Zermelo's Navigation Problem on the Sphere}
\label{sec:zermelo-spherical}

We now modify the above derivations to the case where the ship is travelling on the surface of the Earth - idealized here as a perfect sphere.  To that end, we adopt spherical coordinates $x_1=\theta$ (longitude) and $x_2=\phi$ (latitude)  measured in  units of $\kappa$ radians.  In particular, it may be convenient to take $\kappa = \pi/180$ if we wish to measure things using degrees. The background current  will be given relative to a east-north framing, which we  represent as the following $2\times 2$ matrix
\[ F(\theta,\phi) =
\begin{bmatrix}
  K \cos\theta & 0\\
  0 & K
\end{bmatrix} \]
where $K$ is the conversion scale from the units used to measure $\theta,\ \phi$ and the units used to measure local velocities.  For example, if global position is measured using degrees of arc, and local velocities are measured in kilometres, then letting $R$ be the earth's radius in kilometres ($R \approx 6367$ km), we have $K = \kappa R = \pi R/180\approx 111.1$ kilometres per 1 degree of arc.

With these conventions in place, the velocity over ground of a vehicle moving at a speed of $V$ over water is given as
\begin{align}
\begin{split}
  K\cos(\kappa\phi)\frac{d\theta}{dt}
  &= V\cos(\kappa\alpha)+w_1(\theta,\phi)\\ 
  K\frac{d\phi}{dt} &= V\sin(\kappa\alpha)+w_2(\theta,\phi)
\end{split}
\label{eq:dxdt-spherical}
\end{align}
where $\alpha$ is the ship's heading measured relative to an East-North framing, where $\vec{w}(\theta,\phi) = \langle w_1(\theta,\phi), w_2(\theta,\phi)\rangle$ with $w_1$ being the component of displacement relative to east, and $w_2$ the component of displacement relative to north.

Using the Calculus of Variations once again, now on the sphere, one can show that such a path necessarily obeys the following differential equation,

\begin{align}
\begin{split}
\kappa K
  \frac{d\alpha}{dt}
  &=
    \begin{bmatrix}
      \cos(\kappa\alpha) & \sin(\kappa\alpha)
    \end{bmatrix}
    \begin{bmatrix}
      \sec(\kappa\phi) w_{1,1} & w_{1,2}\\
      \sec(\kappa\phi) w_{2,1} & w_{2,2}
    \end{bmatrix}
    \begin{bmatrix}
\sin(\kappa\alpha)\\-      \cos(\kappa\alpha)
\end{bmatrix} \\
  &\quad -
  \cos(\kappa\alpha)\tan(\kappa\phi)(V+\cos(\kappa\alpha) w_1+
  \sin(\kappa\alpha) w_2)
\end{split}
\label{eq:dthetadt-spherical}
\end{align}

Equation \eqref{eq:dthetadt-spherical} can be justly considered the analogue of the Zermelo differential equation for motion on a sphere. For the sake of completeness, the derivation of this equation is fully explained in Appendix \ref{appendix:zermelo}.


\section{Hybrid Search method}
\label{sec:hybrid}

\acf{HS} is the 3-step algorithm proposed in this paper for solving the Zermelo-problem in either Euclidean or Spherical background. The 3 steps are (i) exploration, (ii) refinement, and (iii) smoothing. The output of the exploration and refinement phases is a piece-wise optimal trajectory that connects a starting location with a desired destination.

In effect, exploration is a shooting method based on the \acf{ZIVP}. The exploration algorithm formulates multiple instances of a \ac{ZIVP} with a given initial position and a fan of directions aimed towards the target. The trajectories are then evolved using  \ac{RK4} numerical solutions to the Zermelo Differential Equation with dynamic termination conditions. The most obvious termination condition is to select the trajectory that minimizes the distance to the target.  In practice, it turns out that a better heuristic is to terminate each trajectory when the difference between the heading angle and the direction to target exceeds a certain pre-set threshold.  The algorithm is greedy, in that a single ``winner'' trajectory is selected from the list of dynamically terminated trajectories.  This selection is performed on the basis of distance to target.

The refinement phase is just a refined version of the exploration algorithm, but this time the fan of initial directions is taken as small deviations from the winner trajectory of the Exploration phases. The tightness of the refinement spread is constrained so that the fan of directions does not exceed the spread between two directions of the exploration phase.  The candidate trajectories are then evolved using the same heuristic as in the exploration phase and a winner is selected based on proximity to the target. The precise details of the exploration and refinement sub-algorithms are detailed in sections \ref{sec:hybrid-exploration} and \ref{sec:hybrid-refinement} respectively.

The third phase consists of smoothing the output of the refinement using the \ac{FMS} algorithm.  This algorithm is a numerical Boundary Value Problem scheme that works by iteratively shifting a given discretized trajectory towards a time-minimizing route.  The approach is based on the discrete Calculus of Variation and can, in principle, be utilized with any given Lagrangian.  In our case, we select the time-minimizing Lagrangian such that the corresponding Euler-Lagrange equations are precisely the Zermelo Differential Equation.   We then discretize the time-minimizing Lagrangian using a pre-selected time-step and begin the iteration with the piece-wise optimal solution generated by the Exploration and Refinement sub-algorithms.  Because the initial trajectory is piece-wise optimal, the overall effect is that of smoothing the sharp turns present in the initial trajectory and converting the piece-wise smooth and piece-wise optimal solution to a smooth, near optimal solution of the Zermelo problem.  The relevant details of the \ac{FMS} algorithm are specified in section \ref{sec:fda}.

\subsection{Exploration step}
\label{sec:hybrid-exploration}

Given a start point $\boldsymbol{x_A} = (x_{A, 1}, x_{A, 1})$, and a goal point $\boldsymbol{x_B} = (x_{B, 1}, x_{B, 2})$, we can first centre a ``search cone'' in the direction of $\Lambda_{A, B}$, following equation \eqref{eq:shooting-angle} (assuming an euclidean space). The amplitude for this cone is $\gamma$. If the vector field was null and we started a trajectory with heading $\alpha = \Lambda_{A, B}$, the vessel would eventually arrive to $\boldsymbol{x_B}$. Thus, by taking this search cone, we are assuming that the optimal route will always point close to the destination and that the vector field will have a small effect on the vessel trajectory. However, this assumption can be relaxed by increasing the amplitude of the cone, $\gamma$ (up to $2\pi$, covering all directions). 

\begin{equation}
    \Lambda_{i, j} = \Lambda(\boldsymbol{x_i}, \boldsymbol{x_j}) = \arctan\left(\frac{x_{j,2} - x_{i,2}}{x_{j, 1} - x_{i, 1}}\right)
    \label{eq:shooting-angle}
\end{equation}

Equation \eqref{eq:shooting-angle} defines the angle $\Lambda_{i, j}$ from point $\boldsymbol{x_i}$ to point $\boldsymbol{x_j}$. This equation is applicable in Euclidean space, and can be generalized to spherical geometry for short distances. However this does not hold for our study as distances between start and end points are significant, so when working in spherical space it is better to replace equation \eqref{eq:shooting-angle} by the following:

\begin{equation}
    \Lambda_{i,j} = \arctan \left( \frac{-c_j \cdot s_i + c_i \cdot s_j}{-\left( c_i \cdot c_j + s_i \cdot s_j \right) \cdot \sin (x_{i, 2}) + \left( c_i^2 +s_i^2 \right) \cdot \sin (x_{j, 2})} \right)
    \label{eq:shooting-angle-spherical}
\end{equation}

where $c = \cos (x_1) \cdot \cos (x_2)$; and $s = \sin (x_1) \cdot \cos (x_2)$.

Next, we generate $N$ initial shooting angles, namely
\[\alpha_n(0) \in \left[ \Lambda_{A, B} - \gamma / 2,\ \Lambda_{A, B} + \gamma / 2 \right].\]
To do so we $N$-sect the ``search cone'' into $\alpha_0, \dots, \alpha_N$, evenly spread across the whole search cone, and use each of these as an initial condition to solve the system of ODE via the \ac{RK4}. We will use these shooting angles to generate $N$ local paths, or trajectories $q_n(t) = ( x_{n, 1}(t), x_{n,2}(t), \alpha_n(t) ),\ n \in [0, N]$.

The $N$ generated trajectories evolve using \ac{RK4}, in iterations of time $\tau > \Delta t$ (where $\Delta t$ is the time step of \ac{RK4}). That means that, after iteration $i$, the routes will have evolved until time $t = i \tau$, and will be defined by the points $q_n(t),\ t \in [0, i\tau]$. We name the first iteration of this exploration step as $i_0$, which will start at $i_0 = 1$ but will be updated in further optimization steps.  After every iteration $i$, each trajectory $n$ is checked individually to assert whether it meets any one of three stopping conditions. If it does, trajectory $n$ is left out of the \ac{RK4} loop and won't evolve further. These three rules are:

\begin{enumerate}
    \item Trajectory $n$ is stopped after iteration $i$ if
    \[D\left( \boldsymbol{x_n}(i\tau), \boldsymbol{x_B} \right) \leq d,\]
    being $D(\boldsymbol{x_a}, \boldsymbol{x_b})$ the distance metric between two points, defined according to the space we are operating on, and $d$ a certain distance threshold. This implies the vessel has reached its goal.
    
    \item Trajectory $n$ is stopped after iteration $i$ if its heading $\alpha_n(i\tau)$ deviates too much from the goal. To assert this, we take point  $\boldsymbol{x_n}(i\tau)$, and compute its angle to $\boldsymbol{x_B}$, named $\Lambda(\boldsymbol{x_n}(i\tau), \boldsymbol{x_B})$, see equations \eqref{eq:shooting-angle} and \eqref{eq:shooting-angle-spherical}. Otherwise, the trajectory keeps evolving while the following condition is met:
    \[\left( \Lambda(\boldsymbol{x_n}(i\tau), \boldsymbol{x_B}) - \gamma_d / 2 \right) \leq \alpha_n(i\tau) \leq \left( \Lambda(\boldsymbol{x_n}(i\tau), \boldsymbol{x_B}) + \gamma_d / 2 \right),\]
    where $\gamma_d$ is the maximum deviation allowed from the goal, typically equal or lower than the search cone $\gamma_d \leq \gamma$. The higher $\gamma_d$, the more exploratory is this method, but it will take more iterations to converge.

    \item Trajectory $n$ is stopped after iteration $i$ if any of its points $\boldsymbol{x_n}(t), t \in [0, i\tau]$ is located in land. In addition to stopping the trajectory, the algorithm discards all the way-points $q_n(t),\ t \geq t_{\text{land}}$, being $\boldsymbol{x_n}(t_{\text{land}})$ the first point located in land. The trajectory $q_n(t),\ t < t_{\text{land}}$ is kept, as it may still be the optimal route and just needs a course correction, that will be done in a later step.
    
\end{enumerate}

One can argue that rule 2 is too strict for small $\gamma_d$, as the vessel can be heading ``wrongly'' for a negligible amount of time before turning ``correctly'' again, and that the resulting route might be optimal. However, when working with real scenarios, the influence of the vector field is small enough to justify that a vessel going in a ``wrong'' direction won't turn ``correctly'' on time to compensate this deviation.

\begin{figure}[!htb]
    \centering
    \begin{subfigure}[b]{0.49\textwidth}
        \centering
        \includegraphics[width=\linewidth]{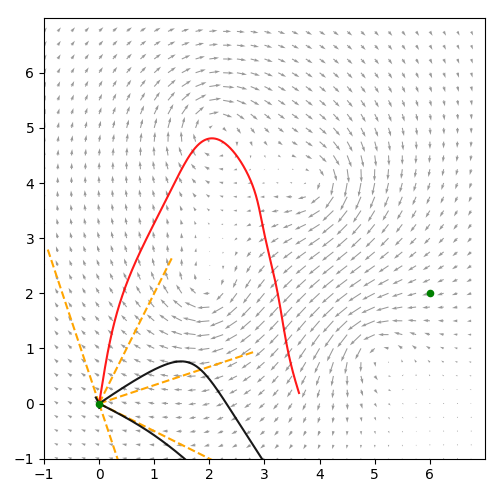}
        \caption{}
        \label{fig:exploration}
    \end{subfigure}
    \begin{subfigure}[b]{0.49\textwidth}
        \centering
		\includegraphics[width=\linewidth]{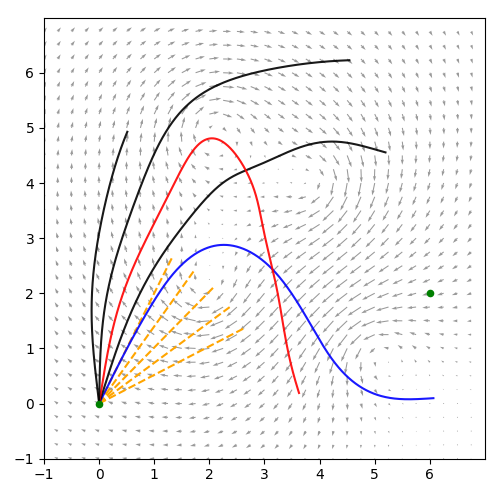}
        \caption{}
        \label{fig:refinement}
	\end{subfigure}
\caption{First two steps of the \acf{HS} method: (a) exploration and (b) refinement. Each trajectory is generated from a different shooting angle (in orange) and evolves using \acf{RK4} method iteratively with $\tau=0.1$, until their heading deviates more than $\gamma_d = \pi/2$ radians from the goal. After all local paths are computed, the one that got closer to the destination is chosen as best (highlighted in the graph). The search cone had an amplitude of $\gamma = \pi$ radians in the exploration step and was centred on the direction of the goal. During refinement, the search cone was centred on the shooting angle of the best route found in the exploration step, and its amplitude is narrower, $\gamma = \pi/5$.}
\end{figure}

Figure \ref{fig:exploration} shows a visualization of this exploration step, highlighting the one which got closest to the goal. \ac{RK4} method ensures that all trajectories are time optimal. After all $N$ trajectories stop, if none of them reached the goal $\boldsymbol{x_B}$ (i.e. none met the 1st stopping rule), we keep the trajectory
\begin{equation}
    m: D\left( \boldsymbol{x_{m}}(i_1\tau), \boldsymbol{x_B} \right) \leq D\left( \boldsymbol{x_{n}}(i_1\tau), \boldsymbol{x_B} \right) \forall n \in [0, N]
    \label{eq:trajectory-best}
\end{equation}
where $i_1$ was the last iteration from \ac{RK4} method. We named this trajectory $m$ as our ``best trajectory'', then move to the refinement step (section \ref{sec:hybrid-refinement}).

\subsection{Refinement step}
\label{sec:hybrid-refinement}

In the exploration step, we assumed that the optimal route should be heading closely towards the goal $\boldsymbol{x_B}$, and evolved trajectories defined by the points

\[q_n(t),\ n \in [0, N],\ t \in [0, i_1\tau].\]

with initial shooting angles $\alpha_n(0) \in \left[ \Lambda_{A, B} - \gamma / 2,\ \Lambda_{A, B} + \gamma / 2 \right]$.

We now generate a narrower search cone, with amplitude $\gamma_b << \gamma$ (for instance, $\gamma_b = \gamma / 5$) and we centre it on $\alpha_{m}$, where $m$ is the ``best trajectory'' from the exploration step. Thus, the newly generated initial shooting angles are evenly spread across

\[\alpha_n(0) \in \left[ \alpha_{m} - \gamma_b / 2,\ \alpha_{m} + \gamma_b /2 \right],\ n \in [0, N]\]

We now redo the exploration step, starting at iteration $i = i_0$. The first time we enter the refinement step, $i_0=1$ but that will be updated soon. The trajectories will stop eventually, at iteration $i_2$. Note that, as there are new trajectories, we now may take a different number of loops than the exploration step, and thus $i_2$ is not guaranteed to be equal to $i_1$. If no trajectory reached the goal $\boldsymbol{x_B}$ (i.e. no trajectory meets the first stopping criteria), we regenerate update the ``best trajectory'' $m$ following equation \eqref{eq:trajectory-best}. The algorithm goes back to the exploration step (section \ref{sec:hybrid-exploration}) using $\boldsymbol{x_{m}}(i_2\tau)$ as the starting point, i.e. $\boldsymbol{x_A} = \boldsymbol{x_{m}}(i_2\tau)$; and starting at iteration $i_0=i_2$.

This loop between exploration-refinement continues until the first stopping rule happens, i.e. one trajectory gets close enough to the destination $\boldsymbol{x_B}$. Figure \ref{fig:optimized} displays one possible result of this process. One issue is apparent: the vessel takes sharp turns in the connections between local paths. This happens because each trajectory (except the last one) is stopped due to deviating from the goal, so the vessel is forced to correct its course by turning sharply to reach its destination.


\subsection{Smoothing step}
\label{sec:fda}

We see that if we use the paths generated by our approach, it's not very ``smooth'', which is not realistic for real world situations. So we want to ``smooth'' it out while still optimizing the cost. Following \citet{Ferraro2021}, we apply the Newton-Jacobi method to the discretized Euler-Lagrange equation to smooth out the path.  Throughout this paper we will refer to this algorithm as \acf{FMS}. Since each local path is already local optimal, the \ac{FMS} algorithm can converge to an optimal solution after a suitable number of iterations.

\begin{figure}[!htb]
\centering
\begin{subfigure}[b]{0.49\textwidth}
    \centering
    \includegraphics[width=\linewidth]{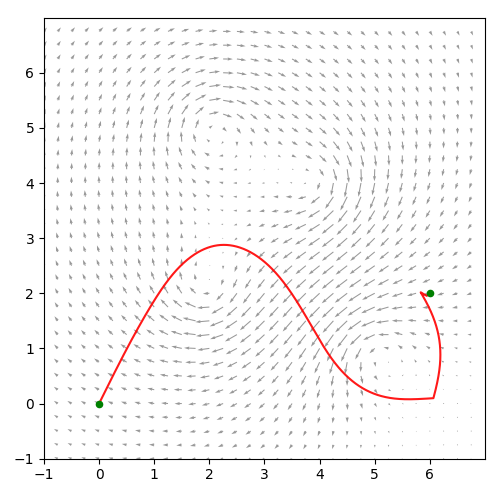}
    \caption{}
    \label{fig:optimized}
\end{subfigure}
\begin{subfigure}[b]{0.49\textwidth}
    \centering
    \includegraphics[width=\linewidth]{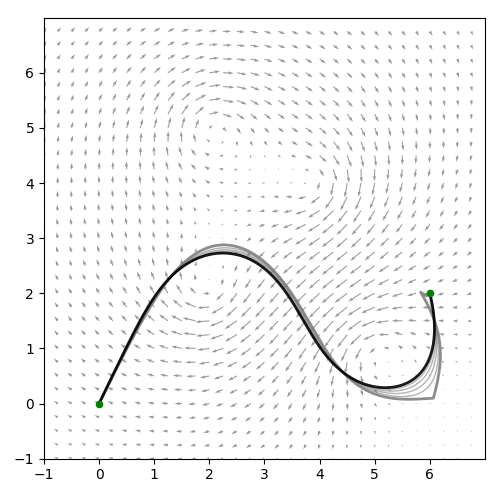}
    \caption{}
    \label{fig:fda}
\end{subfigure}
\caption{(a) Optimized route obtained by alternating the first two steps of \acl{HS} method. The segments are locally optimal (thanks to \acs{RK4}) but are joined by sharp turns. (b) The whole route is then smoothed with \acl{FMS} method for \num{10000} iterations.}
\end{figure}

Let us quickly review the Newton-Jacobi iterative procedure for solving nonlinear equation. Consider an equation of the form $0=f(x)$ where  $f(x)$  is a differentiable function of one variable.  Newton's method proposes that we pick an approximate solution $x=x^{(0)}$ and then solve the linearized system
\[ f(x^{(0)}) + f'(x^{(0)})(x^{(1)}-x^{(0)}) = 0 \]
to obtain an $x^{(1)}$.  If $x^{(0)}$ is sufficiently close to a root of $f(x)=0$, one can show that $|f(x^{(1)})|< |f(x^{(0)})|$ and we can iterate to produce a sequence $x^{(0)},x^{(1)},x^{(2)},\ldots$ by solving, at each stage the linearized system
\[ f(x^{(i)}) + f'(x^{(i)})(x^{(i+1)}-x^{(i)}) = 0.\]

The Newton-Jacobi method generalizes Newton's method to the case of an $n\times n$ system of nonlinear equations $F(q)=0$ where $q$ is a point in $n$-dimensional space and $F$ is a transformation of $n$-dimensional space; i.e., $F(q)=(F_1(q),\ldots, F_n(q))$ is an $n$-vector of functions.   As above, we begin with an initial guess $q_0$ and then construct a sequence of approximate solutions by solving the linearized equations

\[ F(q_i) + DF(q_i)(q_{i+1}-q_i) = 0 \]

for $q_{i+1}$.   Under suitable assumptions, one can show that the sequence $q_0, q_1, q_2,\ldots$ converges to a zero of $F$.

The discrete Euler-Lagrange equations \eqref{DEL}  are a non-linear system of equations of $n\times (N-1)$ equations.  The key idea introduced in  \cite{Ferraro2021} is to apply apply the NJ method iteratively to primitive 3-point trajectories, i.e. trajectory trajectories path consisting of $q_{k-1}, q_{k}, q_{k+1}$. For each such trajectory we freeze $q_{k-1}, q_{k+1}$ and seek for the optimal placement of $q_k$.  This amounts to a solution of the equation

\[   D_{2} L_{d}(q_{k-1},\bar{q}_{k}) + D_{1}
  L_{d}(\bar{q}_{k},q_{k+1}) = 0 \]

for an unknown $\bar{q}_{k}$.  We now apply the NJ method by taking

\[ F(q) =  D_{2} L_{d}(q_{k-1},q) + D_{1}
  L_{d}(q,q_{k+1})  \]

and apply  one iteration of the method to solve the linearized system

\[ F(q_k) + DF(q_k)(q^*_k - q_k) = 0 \]

for the unknown $q^*_k$.   Fully written, the system for $q^*_k$ is then

\[ \begin{split}
 D_2 L_d(q_{k-1},q_k) &+ D_1 L_d(q_k,q_{k+1}) + \\&+ \left(
    D_{22}(q_{k-1},q_k) + D_{11} L_d(q_k,q_{k+1})\right) (q^*_k - q_k) =
  0 
\end{split}\]

We now apply the same one-step iteration to all the primitive trajectories

\[(q_{k-1},q_k, q_{k+1}),\; k=1,\ldots, N-1\]

to obtain a new trajectory $q^*=(q^*_k)_{k=0}^N$ with $q^*_0=q_0$ and $q^*_N = q_N$.  If the initial trajectory $q^{(0)}$ is well chosen, then the iterated sequence of trajectories $q^{(i)},i=0,1,\ldots$ where $q^{(i+1)} = q^{(i)*}$ converges to a solution of the discretized Euler-Lagrange equations \eqref{DEL}.

We will apply the \ac{FMS} algorithm to the Euclidean Zermelo problem after suitably transforming \eqref{eq:langrangian-euclidean} into a non-constrained optimization problem.  It is possible to extend the \ac{FMS} methodology to spherical backgrounds and to constrained optimization, but we do not pursue these directions in the present paper.

We begin by combining \eqref{eq:constraints-euclidean} into the single constraint

\begin{equation}
  \label{xdottdotWV}
 (\dot{x}_1 - w_1 \dot{t})^2+(\dot{x}_2 - w_2 \dot{t})^2 = V^2\dot{t}^2.
\end{equation}

Setting

\begin{align*}
  X&= \sqrt{\dot{x}_1^2+\dot{x}_2^2}\\
  W&= \sqrt{w_1^2 + w_2^2}
\end{align*}

we  rewrite \eqref{xdottdotWV} as the following quadratic equation in $\dot{t}$:

\[ (V^2-W^2)\dot{t}^2  + 2 X W \cos\beta - X^2 = 0, \]

where $\beta$ is the angle between $\dot{x}$ and $w$. The solution gives us the following unconstrained  Lagrangian:

\[ \hat{L} =\dot{t}= \frac{X}{V^2-W^2}\left(-W \cos\beta+
    \sqrt{V^2-W^2\sin^2\beta}\right) \]

As given, the above $\hat{L}$ is not a regular Lagrangian, and the corresponding $\hat{L}_d$ will not give a convergent \ac{FMS} algorithm.  This difficulty can be remedied by observing that $\hat{L}^2$ \textbf{is} regular, and so we take $\hat{L}^2_d$ as the discrete Lagrangian for our implementation of the \ac{FMS} algorithm.

Figure \ref{fig:fda} shows the results of \ac{FMS} after \num{10000} iterations, applied to the route generated at the end of the exploration and refinement loop.


\section{Benchmarks}
\label{sec:benchmarks}

In order to test our optimization approach, we must define a set of benchmarks, containing as many different scenarios as possible. For instance, presence of land between the two ports, strong opposing currents, or highly variable vector fields. Different scenarios can be easily simulated by the use of synthetic benchmarks, which we will comment later. We want, however, to include some real scenarios within our benchmarks. We do not know the best possible route for all benchmarks, but a good point of comparison is the circumnavigation, i.e. the route of minimum distance, also named geodesic when no land is present between the two points. In most realistic scenarios, the optimal routes are found around the geodesic, specially for higher vessel speeds.

\subsection{Synthetic benchmarks}
\label{sec:benchmark-synthetic}

It is important to test algorithms first on synthetic benchmarks for several reasons. For starters, synthetic benchmarks provide a controlled and consistent environment for testing, allowing for more accurate and reliable results. This is particularly useful when evaluating the performance of an algorithm under different conditions, as synthetic benchmarks can be easily manipulated to simulate a wide range of scenarios. Secondly, synthetic benchmarks allow for the testing of algorithms without the need for real-world data, which can be expensive and time-consuming to obtain. This allows for faster and more cost-effective testing and evaluation of algorithms. Third, synthetic benchmarks can be used to test the robustness and reliability of algorithms. By introducing challenges and variations to the synthetic benchmark, it is possible to assess how well an algorithm can handle different situations and environments. This can provide valuable insights into the limitations and potential improvements of the algorithm.

\subsubsection{Circular vector field}
\label{sec:benchmark-circular}

A very simple benchmark we can define is a circular vector field, centred in $(a, b)$. The currents spin clock-wise and have an increasing intensity (defined by $s$) the further one strays from the centre. This is summarized in the following equation:

\[W(x_1,x_2) = \left\langle s \cdot (x_2-b), -s \cdot (x_1-a)\right\rangle\]

In this work, we are using $(a, b)=(-3, -1)$ and a scale factor $s=0.05$, small so that a vessel with unitary velocity can overcome currents near the centre. When testing optimizers we will ask them to develop a path traversing the centre: from $\boldsymbol{x_A}=(3, 2)$ to $\boldsymbol{x_B}=(-7, 2)$. Algorithms are expected to follow the direction of favourable currents.

\subsubsection{Four vortices}
\label{sec:benchmark-fourvortices}

The next synthetic benchmark we will test is the one appearing in \citet{Ferraro2021}, called ``four vortices''. This vector field is defined by the following equation:

\[W(x_1,x_2) = s \cdot \left( -R_{2, 2} - R_{4,4} -R_{2,5} + R_{5,1} \right),\]

where each vortex is expressed as

\[
R_{a,b}(x_1, x_2) = \frac{1}{3\left( (x_1 - a)^2 + (x_2 - b)^2\right) + 1}
\begin{bmatrix}
    -(x_2 - b) \\ x_1 - a
\end{bmatrix}.
\]

The authors explained that the scale factor $s=1.7$ is chosen so that the maximum value of $|W|$ is almost 1. As we are testing these synthetic benchmarks using vessel with unitary velocities, we respect this factor. When testing optimizers we will ask them to develop a path from $\boldsymbol{x_A}=(0, 0)$ to $\boldsymbol{x_B}=(6, 2)$.

\subsection{Real benchmarks}
\label{sec:benchmark-real}

Oceanographic data for real case scenarios was downloaded from Copernicus Marine Environment Monitoring Service \cite{Copernicus2019}. Copernicus offers APIs and a Python client to facilitate and automate data downloads, which are typically stored in NetCDF format.

\begin{figure}[!htb]
\centering
\begin{subfigure}{.5\textwidth}
  \centering
  \includegraphics[width=\linewidth]{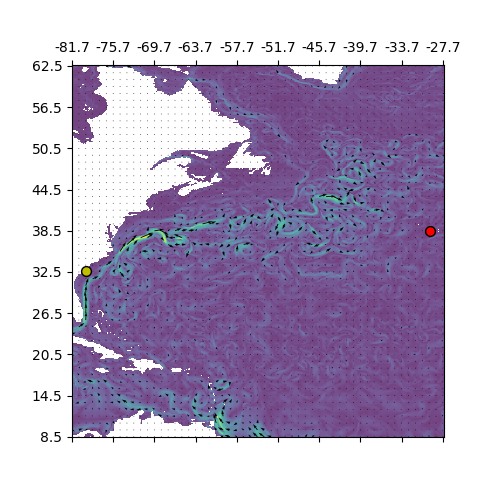}
  \caption{Charleston-Azores}
  \label{fig:vectorfield-charleston}
\end{subfigure}%
\begin{subfigure}{.5\textwidth}
  \centering
  \includegraphics[width=\linewidth]{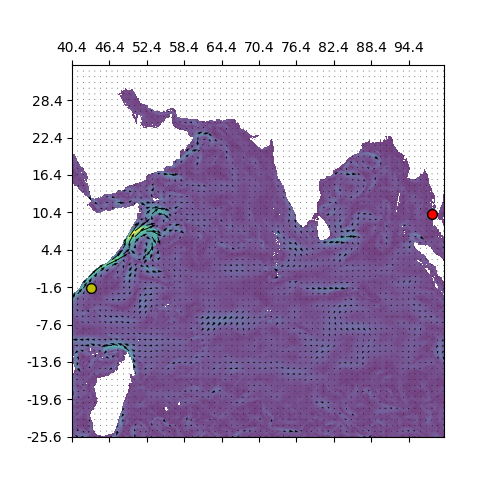}
  \caption{Somalia-Myanmar}
  \label{fig:vectorfield-somalia}
\end{subfigure}%
\vskip\baselineskip
\begin{subfigure}{.5\textwidth}
  \centering
  \includegraphics[width=\linewidth]{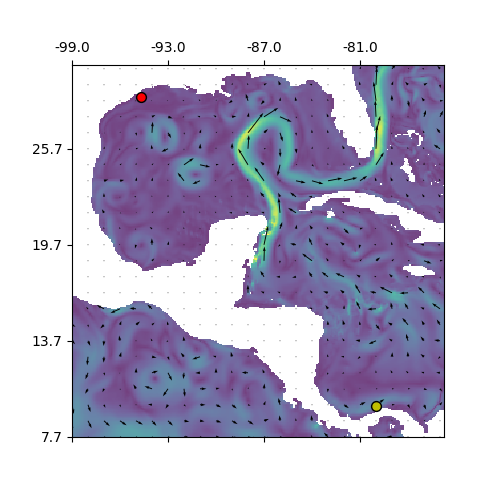}
  \caption{Panama-Houston}
  \label{fig:vectorfield-panama}
\end{subfigure}%
\begin{subfigure}{.5\textwidth}
  \centering
  \includegraphics[width=\linewidth]{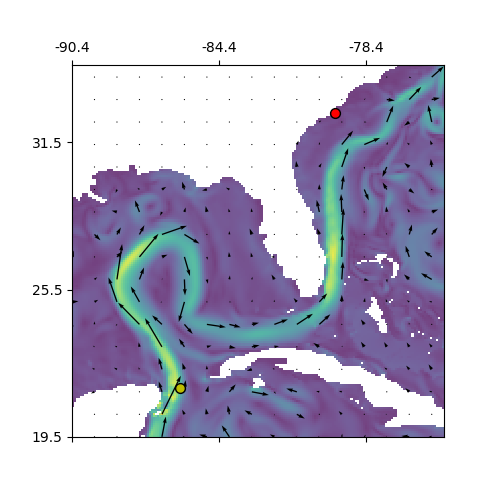}
  \caption{Cancun-Charleston}
  \label{fig:vectorfield-cancun}
\end{subfigure}%
\caption{Real vector fields. Yellow point marks the starting position and red is the goal. The ocean currents are coloured by intensity, the fastest being represented with brighter (greener) colours.}
\label{fig:vectorfield-real}
\end{figure}

As a first example, we consider a journey from Charleston ($32.7\degree$N $79.7\degree$W) to the Azores islands ($38.5\degree$N $29.5\degree$W) during spring, using data specifically from May 25th, 2022. This represents the simplest real-world scenario as the trajectory is largely over open ocean, and ocean currents are relatively calm in this region of the Atlantic. However, a favourable current flows northeastward near the departure point (see Figure~\ref{fig:vectorfield-charleston}), which could be exploited by our algorithm to save time.

The second example is a journey from Somalia ($1.66\degree$S, $42.39\degree$E) to Myanmar ($10.21\degree$N, $98.14\degree$E), traversing the Indian Ocean during summer, with data specifically obtained from the 1st of July 2022. This scenario presents a greater challenge than the first example, as there are several islands along the way that the vessel must avoid in order to reach its destination safely, as shown in Figure~\ref{fig:vectorfield-somalia}. Our goal in using this benchmark is to test the algorithm's ability to avoid land while also making use of the vector field.

The third example involves a journey from Panama ($9.7\degree$N, $80.0\degree$W) to Houston ($29.0\degree$N, $94.7\degree$W), traversing the Caribbean and the Gulf of Mexico. The data used for this example is the same as in the first example, from the 25th of May 2022. In this case, we are testing the algorithm's ability to avoid large land masses, while also utilizing the Gulf Stream, which is clearly visible in Figure~\ref{fig:vectorfield-panama}.

The final example is a journey from Cancun ($21.5\degree$N, $86.0\degree$W) to Charleston ($32.7\degree$N, $79.7\degree$W), once again traversing the Gulf of Mexico and reaching the Atlantic Ocean. We again use the same data as in the previous example, from the 25th of May 2022. This scenario presents the greatest challenge of all, as the vessel must navigate through a narrow waterway connecting two large land masses (Florida and Cuba), as shown in Figure~\ref{fig:vectorfield-cancun}. In this case, we will test which parameters work best for the algorithm to avoid large land masses.


\section{Results}
\label{sec:results}

We run the \acf{HS} algorithm in all the benchmarks mentioned in section~\ref{sec:benchmarks}, aiming to find the route that takes the least amount of time. The results of the algorithm are shown twice: first after applying \ac{HS} to solve the \acf{ZIVP}, and then after smoothing the previous result with the \ac{FMS} algorithm. To have a point of comparison, we also show the time elapsed by the route of minimum distance. For the synthetic vector fields, operating in Euclidean geometry, the minimum distance is the straight line. For the real vector fields, the minimum distance is the geodesic in the absence of land, otherwise it is called the circumnavigation route.


\subsection{Synthetic benchmarks}
\label{sec:results-synthetic}

To solve the synthetic vector fields for a vessel of unitary velocity on Euclidean geometry, \acf{HS} was run using a time step of $\Delta t = 0.01$ and checking the stopping criteria every $\tau = 0.1$. There were twenty one trajectories being tested by the \ac{HS}, their initial shootings evenly spread across a cone of amplitude $\gamma = \pi$ centred on the direction to the goal. The trajectories would stop if their heading deviated more than $\gamma_d = \pi/2$, or when they got close to the goal (at least $d = 0.1$). Once the \ac{HS} guessed and optimal route, it would be smoothed by the \ac{FMS} algorithm during \num{10000} iterations.

\begin{figure}[!htb]
\centering
\begin{subfigure}{.5\textwidth}
  \centering
  \includegraphics[width=\linewidth]{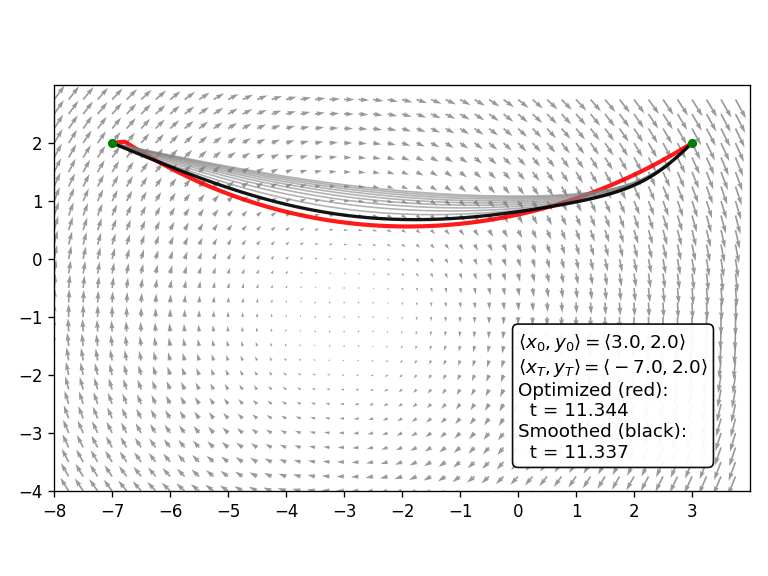}
  \caption{Circular}
  \label{fig:results-circular}
\end{subfigure}%
\begin{subfigure}{.5\textwidth}
  \centering
  \includegraphics[width=\linewidth]{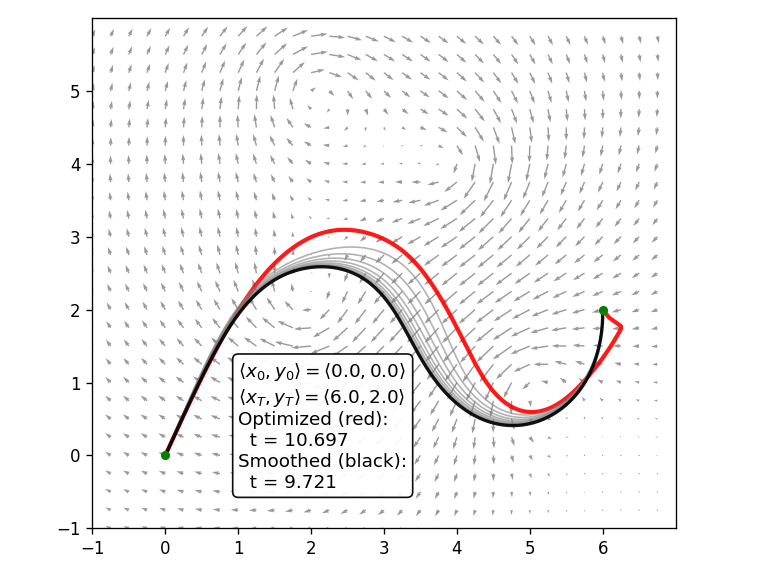}
  \caption{Four vortices}
  \label{fig:results-four-vortices}
\end{subfigure}
\caption{Results on the synthetic vector fields, sailing at unit speed.}
\label{fig:results-synthetic}
\end{figure}

The travel times obtained by our method in the two synthetic benchmarks are shown in Table~\ref{tab:results-synthetic}. In addition we include the optimized route for the four vortices vector field in Figure~\ref{fig:fda}, which coincidentally is the best route found by the original designers of the benchmark \cite{Ferraro2021}, proving that the \ac{HS} method is well-suited to give initial guesses that can be smoothed with \ac{FMS}.

\begin{table}[!htb]
    \centering
    \begin{tabular}{llrH}
        \toprule
        Vector field & Method & Time & Comp. time\\
        & & & (min) \\
        \midrule
        Circular & Min. dist. & 11.93 & - \\
        & \ac{HS} & 10.56 & 11 \\
        [0.5em]
        Four Vortices & Min. dist. & 30.44 & -  \\
        & \ac{HS} & 9.72 & 12 \\
        \bottomrule
    \end{tabular}
    \caption{Results on synthetic vector fields with unitary velocity, comparing the route of minimum distance with the output from \acf{HS} method.}
    \label{tab:results-synthetic}
\end{table}

In both benchmarks, the \ac{HS} method is able to find a route that takes less time than travelling the minimum distance. Improvements are great in the four vortices vector field, due to the currents having a velocity close to the vessel and for that reason having a bigger effect. The circular vector field has currents with lower speeds, and thus the optimized route is not that different from the route of minimum distance.


\subsection{Real benchmarks}
\label{sec:results-real}

To solve the real vector fields for a vessel of different velocities on spherical geometry (approximating the Earth's radius as $6367.449$ km), \acf{HS} was run using a time step of $\Delta t = 600$ s (10 minutes) and checking the stopping criteria every $\tau = 7200$ s (2 hours). There were twenty one trajectories being tested by the \ac{HS}, their initial shootings evenly spread across a cone of amplitude $\gamma = \pi$ ($180\degree$) centred on the direction to the goal. The trajectories would stop if their heading deviated more than $\gamma_d = \pi/2$ ($90\degree$), or when they got close to the goal (at least $d = 10$ km). Once the \ac{HS} guessed and optimal route, it would be smoothed by \acf{FMS} during \num{2000} iterations.

\begin{table}[!htb]
	\centering
	\begin{tabular}{lrlrrH}
		\toprule
		Benchmark  & Speed & Method     & Travel time & Distance & Comp. time \\
		           & (m/s) &            &         (h) &     (km) &      (min) \\ \midrule
		Charleston &     3 & Min. dist. &       416.4 &   4392.5 &          - \\
		Azores     &       & \acl{HS}    &       389.2 &   4426.4 &        \\
		[0.25em]   &     6 & Min. dist. &       207.9 &   4392.5 &          - \\
		           &       & \acl{HS}    &       202.0 &   4423.3 &        \\
		[0.25em]   &    10 & Min. dist. &       124.9 &   4392.5 &          - \\
		           &       & \acl{HS}    &       123.3 &   4418.3 &        \\ \midrule
		Somalia    &     3 & Min. dist. &       552.2 &   6159.8 &          - \\
		Myanmar    &       & \acl{HS}    &       528.3 &   6196.9 &        \\
		[0.25em]   &     6 & Min. dist. &       280.0 &   6159.8 &          - \\
		           &       & \acl{HS}    &       274.8 &   6190.4 &        \\
		[0.25em]   &    10 & Min. dist. &       169.3 &   6159.8 &          - \\
		           &       & \acl{HS}    &       167.8 &   6190.4 &        \\ \midrule
        Panama     &     3 & Min. dist. &       242.5 &   2672.7 &          - \\
		Houston    &       & \acl{HS}    &       230.8 &   2719.7 &        \\
		[0.25em]   &     6 & Min. dist. &       124.0 &   2672.7 &          - \\
		           &       & \acl{HS}    &       120.9 &   2700.8 &        \\
		[0.25em]   &    10 & Min. dist. &       74.2 &   2672.7 &          - \\
		           &       & \acl{HS}    &       74.0 &   2716.6 &        \\ \midrule
        Cancun     &     3 & Min. dist. &    130.7  & 1350.4   &        - \\
		Charleston &       & \acl{HS}    &     119.8 &   1381.9 &        \\
		[0.25em]   &     6 & Min. dist. &      70.0 &   1350.4 &          - \\
		           &       & \acl{HS}    &       65.6 &   1354.7 &        \\
		[0.25em]   &    10 & Min. dist. &      42.9 &   1350.4 &          - \\
		           &       & \acl{HS}    &       40.8 &   1365.8 &        \\ \bottomrule
	\end{tabular}
	\caption{Results on the real benchmarks, comparing the route of minimum distance (circumnavigation) with the output from \acf{HS} method.}
	\label{tab:results-real}
\end{table}

\begin{figure}[!htb]
\centering
\begin{subfigure}{.5\textwidth}
  \centering
  \includegraphics[width=\linewidth]{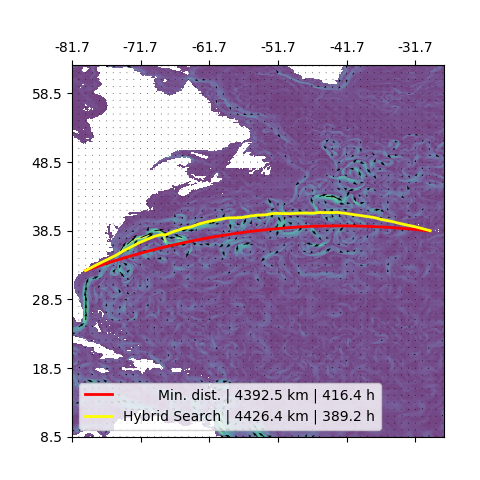}
  \caption{Charleston - Azores}
  \label{fig:results-charleston}
\end{subfigure}%
\begin{subfigure}{.5\textwidth}
  \centering
  \includegraphics[width=\linewidth]{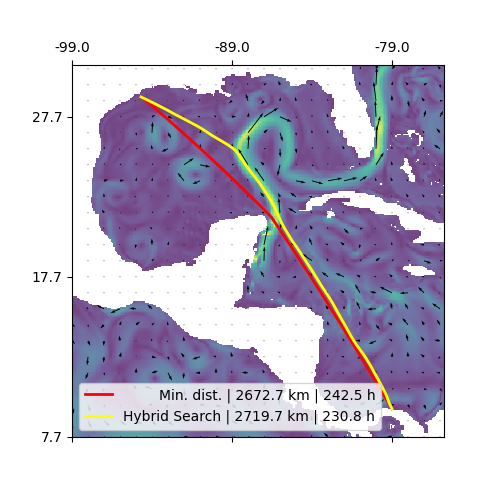}
  \caption{Panama - Houston}
  \label{fig:results-panama}
\end{subfigure}
\caption{Results on the real vector fields, sailing at 3 m/s.}
\label{fig:results-real}
\end{figure}

Figure~\ref{fig:results-real} shows how the optimization method meets our expectations: the route deviates to the north and follow a strong current. This makes the vessel cover more distance but in turn reduces the amount of travel time. The effect of this current is relevant enough even for the fastest vessel speeds.

\section{Summary and Discussion}
\label{sec:discussion}

This paper introduced a new method to solve the \acf{ZIVP}, named the \acf{HS} method. The \ac{HS} method employs a technique of shooting different guesses centred around the general direction towards the goal and simulates the effect of the vector field on these trajectories by applying the fourth-order \acf{RK4} method. By alternating between an exploratory phase and a refinement step, and following some stop rules to eliminate sub-optimal guesses, the \ac{HS} method outputs a chain of trajectories that connect the departure and goal points, with each trajectory being locally optimal due to the \ac{RK4} method. The solution from these initial steps can then be used as input for \acf{FMS}, which returns a locally optimal solution. The advantages of this whole approach are plenty: \ac{FMS} converges faster to a solution when given the locally optimal trajectories from \ac{RK4}, and the sharp turns inherent to the \ac{RK4} method are smoothed by \ac{FMS}.

While the \ac{HS} method does not guarantee finding the global optimum, its search parameters can be tuned to emphasize its exploratory nature, for instance by opening the search cone or allowing bigger deviations from the goal. The more exploration that is allowed, the better local optima can be found, in exchange for increased computation time. Furthermore, the algorithm can be made less greedy by keeping the best $N$ trajectories when moving between exploration and refinement steps, instead of focusing just on the one that gets closest to the goal. This is a future line of work to improve the \ac{HS}.

The \ac{HS} algorithm is easily adapted to a spherical space, making it well-suited for its application in real maritime routing problems. Results also show how by fine-tuning the parameters of the algorithm, the \ac{HS} method finds an optimal route while avoiding land masses. This ability of the \ac{HS} method can be extrapolated not only to obstacle avoidance but also to circumnavigate dangerous areas or no-sail zones. In addition, by changing the time problem to a realistic consumption model through the use of calculus of variations and modifying the Hamiltonian and initial value problem, our algorithm can be easily adapted to a variety of situations. This makes our approach well-suited for real-world applications in weather routing.

Our results demonstrate that the \acf{HS} method consistently outperforms the route of minimum distance in terms of travel time and distance for navigating vector fields with complex structures. This is particularly evident in the case of the synthetic vector fields, where the \ac{HS} method achieves a significant reduction in travel time compared to the reference route. In fact, the travel time reduction achieved by the \ac{HS} method ranges from $11\%$ to $68\%$ depending on the complexity of the vector field, as shown in Table \ref{tab:results-synthetic}. These results indicate that the \ac{HS} method is a promising approach for optimizing paths in various types of vector fields.

In addition to the synthetic vector fields, we also tested the \ac{HS} method on four real-world benchmarks, all of which represent common maritime routes. The results, presented in Table \ref{tab:results-real}, demonstrate that the \ac{HS} method consistently outperforms the route of minimum distance in terms of travel time for these benchmarks. Specifically, the \ac{HS} method achieves a travel time reduction of up to $6.5\%$, reducing fuel consumption up to $12.6\%$, when we consider that the fuel consumed typically scales quadratically with travel time \cite{Harvald1992, Bialystocki2016}. Although the reduction achieved in these real-world benchmarks is relatively small compared to the synthetic benchmarks, the \ac{HS} method still shows its potential to improve the efficiency of sailing in real-world scenarios.

In conclusion, the proposed \ac{HS} method provides a powerful tool for solving the \ac{ZIVP} and has shown promising results in real-world applications. The ability to fine-tune the parameters of the algorithm enables it to explore a range of possibilities, providing a flexible and adaptive approach to routing problem-solving. 

\section*{Acknowledgements}

 The research of RM, DP and DGU is supported by the BBVA Foundation via the project ``Mathematical optimization for a more efficient, safer and decarbonized maritime transport''. In addition, the research of DGU is supported in part by the spanish Agencia Estatal de Investigaci\'on under grants PID2021-122154NB-I00 and TED2021-129455B-I00. The authors would like to thank the MITACS Accelerate program that enabled the 3 months visit of DP to Dalhousie University in the summer of 2022, where this work was initiated, and the visit of LB to Madrid in the summer of 2023, where the work was completed.


\bibliographystyle{apalike}
\bibliography{references}

\newpage
\appendix

\section{Derivation of Zermelo's equations}
\label{appendix:zermelo}

\noindent\textbf{A.1. Zermelo's Navigation Problem on the plane}\\

We are dealing here with a constrained optimization problem whose Lagrangian function has the form
\begin{equation}
  \label{eq:langrangian-euclidean}
  L = \dot{t} + \lambda_1( \dot{x}_1 - (V\cos\alpha+w_1)\dot{t}) +
  \lambda_2(\dot{x}_2- (V\sin\alpha + w_2)\dot{t}). 
\end{equation}
The goal is to find trajectories $x(s),\ \dot{x}(s)=x'(s),\ t(s),\ \dot{t}(s)=t'(s)>0,\ \alpha(s)$ with fixed end-points that minimize  $t(s_1)-t(s_0)=\int_{s_0}^{s_1} L ds$, and obey constraints

\begin{equation}
\begin{aligned}
  \dot{x}_1 &= (V\cos\alpha+w_1)\dot{t} \\
  \dot{x}_2 &=  (V\sin\alpha + w_2)\dot{t}
\end{aligned}
\label{eq:constraints-euclidean}
\end{equation}

The quantities $\lambda_1, \lambda_2$ are known as Lagrange multipliers. As we now show, their form is determined by the Euler-Lagrange equations associated with the above Lagrangian, namely
\begin{align}
  \label{eq:EL1}
  &\frac{d  L_{\dot{t}}}{ds} = 0\\
  \label{eq:EL2}
  &L_{x_i} - \frac{d  L_{\dot{x}_i}}{ds} = 0\quad i=1,2 \\
  \label{eq:EL3}
  &L_\alpha =  0,
\end{align}
Equation \eqref{eq:EL1} gives
\[
  \frac{d}{ds}  \left(\lambda_1 (V\cos\alpha+w_1) +
  \lambda_2(V\sin\alpha+w_2)\right)=0
\] which implies that
\begin{equation}
  \label{eq:lam1}
  \lambda_1 (V\cos\alpha+w_1) +  \lambda_2(V\sin\alpha+w_2) = C
\end{equation}
where $C\ne 0$ is a constant. Equation \eqref{eq:EL3} gives
\begin{equation}
  \label{eq:lam2}
 \lambda_1 \sin\alpha - \lambda_2 \cos\alpha = 0.
\end{equation}
Together, \eqref{eq:lam1} \eqref{eq:lam2} determine the form of the Lagrange multipliers, namely:
\begin{align}
  \lambda_1 &= \frac{ C\cos\alpha}{V+ w_1 \cos\alpha + w_2 \sin\alpha}\\
  \lambda_2 &= \frac{ C\sin\alpha}{V+ w_1 \cos\alpha + w_2 \sin\alpha}
\end{align}

Going forward, we re-parameterize all curves with respect to time $t$ so that
\[ \frac{d}{dt} =  \frac1{\dot{t}} \frac{d}{ds}.\]
E-L equations \eqref{eq:EL2} give the dynamics of the Lagrange multipliers, namely
\begin{align}
  \frac{d\lambda_1}{dt}
  &= -\lambda_1 w_{1,1}- \lambda_2 w_{2,1}\\
  \frac{d\lambda_2}{dt}
  &= -\lambda_1 w_{1,2}- \lambda_2 w_{2,2}
\end{align}
Rewriting \eqref{eq:lam2} as
\[ \tan \alpha = \frac{\lambda_2}{\lambda_1}, \]
and taking derivatives, gives
  \begin{align}
    \nonumber
    \sec^2(\alpha)\frac{d\alpha}{dt}
    &= \frac{d}{dt}\left(
      \frac{\lambda_2}{\lambda_1} \right)\\     \nonumber
    \left(\frac{\lambda_1^2 + \lambda_2^2}{\lambda_1^2}\right)\frac{d\alpha}{dt}
    &=  \frac{1}{\lambda_1^2}\left( -\lambda_2 \frac{d\lambda_1}{dt}  +
      \lambda_1 
      \frac{d\lambda_2}{dt}\right) \\     \nonumber
    \frac{d\alpha}{dt}&= \frac{\lambda_2^2   w_{2,1}+ \lambda_1 \lambda_2
                        (w_{1,1}-w_{2,2})-\lambda_1 ^2 w_{1,2}}
                        {\lambda_1^2+\lambda_2^2} \\
    \frac{d\alpha}{dt}
    &= \sin^2(\alpha)\, w_{2,1}+ \sin(\alpha)\cos(\alpha)\, (w_{1,1}-w_{2,2}) -
      \cos^2(\alpha)\, w_{1,2}
  \end{align}

\noindent\textbf{A.2. Zermelo's Navigation Problem on the sphere}\\

The modified Lagrangian takes the form
\begin{align*}
    L = \dot{t} &+ \lambda_1\left(  \dot{\theta} -K^{-1}
   \sec(\kappa\phi) \left(V\cos(\kappa\alpha)+w_1 \right) \dot{t} \right)
   \\ &+ \lambda_2 \left( \dot{\phi}-K^{-1} \left( V\sin(\kappa\alpha) + w_2 \right)\dot{t} \right)
\end{align*}

The E-L equations \eqref{eq:EL2} now read
\begin{align}
  \label{eq:EL2curved1}
  K\frac{d\lambda_1}{dt}
  &= -\sec(\kappa\phi)\lambda_1 w_{1,1}-\lambda_2
    w_{2,1}\\ 
  K\frac{d\lambda_2}{dt}
  &= - \lambda_1
    \kappa\sec(\kappa\phi)\tan(\kappa\phi)(V\cos(\kappa\alpha)+w_1)
    -\lambda_1\sec(\kappa\phi) 
    w_{1,2}- \lambda_2 w_{2,2} 
\end{align}

In the current setting \eqref{eq:EL3} gives
\begin{align*}
\tan(\kappa\alpha) &= \frac{\lambda_2}{\lambda_1}  \cos(\kappa\phi)
\end{align*}

Taking $d/dt$ yields
\begin{align}
\nonumber
  \kappa\sec^2(\kappa\alpha)\frac{d\alpha}{dt}    
  &=  \frac{\cos(\kappa\phi)}{\lambda_1^2}\left( -\lambda_2
    \frac{d\lambda_1}{dt}  +
    \lambda_1
    \frac{d\lambda_2}{dt}\right)\\
    \nonumber &\quad 
    - \frac{\kappa}{K}\tan(\kappa\alpha)\tan(\kappa\phi) 
    (V\sin(\kappa\alpha)+w_2)\\
\nonumber
  \kappa K \sec^2(\kappa\alpha)
  \frac{d\alpha}{dt} 
  &=
    \frac{\lambda_2}{\lambda_1}w_{1,1}+
    \frac{\lambda_2^2}{\lambda_1^2} \cos(\kappa\phi)w_{2,1}
    - w_{1,2}-
    \frac{\lambda_2}{\lambda_1} \cos(\kappa\phi)w_{2,2}\\
    \nonumber &\quad - \kappa\tan(\kappa\phi)(V\cos(\kappa\alpha)+w_1)\\
    \nonumber &\qquad - \kappa\tan(\kappa\alpha)\tan(\kappa\phi) (V\sin(\kappa\alpha)+w_2)\\
\nonumber
  \kappa K \sec^2(\kappa\alpha)
  \frac{d\alpha}{dt} 
  &= \sec(\kappa\phi)\tan(\kappa\alpha)w_{1,1} + \sec(\kappa\phi)\tan^2(\kappa\alpha) w_{2,1} - w_{1,2}\\
    \nonumber &\quad - \tan(\kappa\alpha)w_{2,2} - \tan(\kappa\phi)(V\cos(\kappa\alpha)+w_1)\\
    \nonumber &\qquad - \kappa\tan(\kappa\alpha)\tan(\kappa\phi) (V\sin(\kappa\alpha)+w_2)\\
\nonumber
  \kappa K
  \frac{d\alpha}{dt}
  &=
    \begin{bmatrix}
      \cos(\kappa\alpha) & \sin(\kappa\alpha)
    \end{bmatrix}
    \begin{bmatrix}
      \sec(\kappa\phi) w_{1,1} & w_{1,2}\\
      \sec(\kappa\phi) w_{2,1} & w_{2,2}
    \end{bmatrix}
    \begin{bmatrix}
\sin(\kappa\alpha)\\-      \cos(\kappa\alpha)
\end{bmatrix} \\
  &\quad -
  \cos(\kappa\alpha)\tan(\kappa\phi)(V+\cos(\kappa\alpha) w_1+
  \sin(\kappa\alpha) w_2)
\end{align}


\section{Euler-Lagrange equations}
\label{appendix:euler-lagrange}

\noindent\textbf{B.1. Continuous Euler-Lagrange equations}\\

Define an action functional along a curve $q(t)$ in $n$-dimensional
space with fixed end points as follows,
\begin{equation} \label{functional} J(q(t)) = \int_{a}^{b} L(t, q(t),
  \dot{q}(t)) dt, \quad q(a)=\alpha, \quad q(b) = \beta.
\end{equation}
The function $L(t, q(t), \dot{q}(t))$ is called the Lagrangian of the
optimization problem. The classical problem in the Calculus of
Variations is to to minimize $J$ by subjecting $q(t)$ to suitable
constraints.

A necessary condition for minimization is that the variation
$\delta J$ vanishes for all possible variations of the trajectory
$\delta q = \epsilon \phi$, where $\phi(t)$ vanishes at the endpoints,
and $\epsilon$ is the variational parameter.  From the functional
\eqref{functional}, define
\begin{equation*}
h(\epsilon) = J(q+\epsilon \phi) = \int_{a}^{b}L(t, q(t)+\epsilon \phi(t), \dot{q}(t) + \epsilon \dot{\phi}(t)) dt.
\end{equation*}
Now differentiate and use the smoothness of $L$ to interchange the derivative and the integral to get
\begin{align*}
h'(\epsilon) &= \frac{d}{d \epsilon} J(q + \epsilon \phi)
= \int_{a}^{b} \frac{d}{d \epsilon} L\left(t, q(t)+\epsilon \phi(t), \dot{q}(t) + \epsilon \dot{\phi}(t)\right) dt \\
&= \int_{a}^{b} \phi(t) \Bigg[ \frac{\partial L}{\partial q} \left(t, q(t) + \epsilon \phi(t), \dot{q}(t) + \epsilon \dot{\phi}(t) \right) \\
&\qquad\qquad\quad  + \dot{\phi}(t) \frac{\partial L}{\partial \dot{q}}\left(t, q(t) + \epsilon \phi(t), \dot{q}(t) + \epsilon \dot{\phi}(t)\right) \Bigg] dt.
\end{align*}
Now setting $\epsilon = 0$ and using our definition of the variational derivative yields
\begin{equation} \label{1st variation}
\delta J(q)(\phi) = \int_{a}^{b} \left[\phi(t)\frac{\partial L}{\partial q}(t, q, \dot{q}) + \dot{\phi}(t)\frac{\partial L}{\partial \dot{q}} (t, q, \dot{q})\right]dt.
\end{equation}
This functional is known as the \textit{first} \textit{variation} of
$J$. In order to obtain an explicit formula for $\delta J$, we need
the integral on the right side of the above equation to be linear in
$\phi(t)$.  We can accomplish this via integration by parts.
\begin{gather*}
  \int_{a}^{b} \dot{\phi}(t) \frac{\partial L}{\partial \dot{q}}(t, q,
  \dot{q})dt = \left[\phi(t)\frac{\partial L}{\partial \dot{q}}(t,
    q(t), \dot{q}(t)) \right]^{t=b}_{t=a} - \int_{a}^{b} \phi(t) \frac{d}{d
    t}\left(\frac{\partial L}{\partial \dot{q}}(t,q,\dot{q}) \right)dt
\end{gather*}
Since $\phi(b)=\phi(a)=0$, by assumption, we obtain the following
formula for the first variation:
 \begin{equation*}
   \delta J(q)(\phi) = \int_{a}^{b} \left[\frac{\partial L}{\partial q} - \frac{d}{dt}\frac{\partial L}{\partial \dot{q}} \right]\phi(t) dt.
 \end{equation*}
 Therefore, in order for $\delta J(\phi)$ to vanish for all $\phi$,
 the critical trajectory $q(t)$ must satisfy the
 \textit{Euler-Lagrange equations}
\begin{equation} \label{EL eq}
\frac{\partial L}{\partial q} - \frac{d}{dt}\frac{\partial L}{\partial \dot{q}} = 0.
\end{equation}

\noindent\textbf{B.2. Discrete Euler-Lagrange equations}\\

Now consider two positions: $q_{0}$ and $q_{1}$, and a time step
$h>0$. We discretize a continuous Lagrangian $L(q,\dot{q})$ by  assuming
that $q_1,q_0$ are close together so that $\dot{q}$ can be
approximated by $(q_1-q_0)/h$.  This allows us to define the following
discrete Lagrangian
\[ L_d(q_0, q_1;h):= \frac{h}{2}\left(
    L\left(q_0,\frac{q_1-q_0}h\right)+
    L\left(q_1,\frac{q_1-q_0}h\right)\right),\]
which approximates the action integral along
a straight trajectory from $q_0$ to $q_1$.  In the discrete Calculus of
Variations, we replace a  continuous curve $q(t)$ with a piece-wise
linear curve determined by a sequence of points
$\{q_{k}\}_{k=0}^{N}$  with $h$ units of time required to go from
$q_k$ to $q_{k+1}$.   We will now calculate the discrete
action over this sequence by summing the discrete Lagrangian.
\[ J_d = \sum_{k=0}^{N-1}L_{d}(q_{k},q_{k+1};h).\] We now vary the
trajectory by $d q = \{ d q_k \}_{k=0}^{N}$ with $d q_0 = d q_N=0$ in
order to fix the boundary points $q_0, q_N$. Note that we use $dq$
rather than $\epsilon \phi$ to describe the variation because the
discretized system has finite degrees of freedom. The variation of the
discrete action can now be given as
\begin{align*}
  d J_d &= \sum_{j=1}^{N-1}\frac{\partial }{\partial
          \boldsymbol{x_j}}\left(\sum_{k=0}^{N-1}  L_d(q_k,q_{k+1};h) \right) dq_j\\ 
        &= \sum_{k=0}^{N-1}\left[D_{1} L_{d}(q_{k},q_{k+1};h) d q_{k} +
          D_{2} L_{d}(q_{k},q_{k+1};h) d q_{k+1} \right]
\end{align*}
Recall that each $\boldsymbol{x_j}=(q_{j1},\ldots, q_{jn})$ is a point in
$n$-dimensional space, so that $\partial/\partial \boldsymbol{x_j},D_1,D_2$ are
actually $n$-vectors of partial derivative operators.  Rearranging the
above sum (this corresponds to the integration by parts step in the
continuous case) we obtain
\[ d J_d = \sum_{k=1}^{N-1}\left[D_{2} L_{d}(q_{k-1},q_{k};h) + D_{1}
    L_{d}(q_{k},q_{k+1};h)\right]d q_{k}.  \] If we require that the
variation of the action is 0 for all $d q_k$, then we obtain the
discrete Euler-Lagrange equations
\begin{equation}
  \label{DEL}
  D_{2} L_{d}(q_{k-1},q_{k};h) + D_{1} L_{d}(q_{k},q_{k+1};h) = 0,\quad
  k=1,\ldots, N-1.
\end{equation}

\end{document}